\documentclass[fleqn,10pt]{wlscirep_JMZ} 

\usepackage{filecontents}
\usepackage{mathrsfs,amsmath}
\usepackage{color,colortbl}
\usepackage{bm}
\usepackage{multirow}
\usepackage{hyperref}
\usepackage{gensymb} 

\definecolor{maroon}{cmyk}{0,0.87,0.68,0.32}

\title{Ionic liquids confined in 1D CNT membranes: gigantic ionic conductivity}

\author[1,2]{Quentin Berrod} 
\author[1]{Patrick Judeinstein}
\author[2]{Yanbao Fu}
\author[2]{Vincent S. Battaglia}
\author[3]{Adeline Fournier}
\author[3]{Jean Dijon}
\author[1,*]{Jean-Marc Zanotti}

\affil[1]{Laboratoire L\'eon Brillouin, CEA, CNRS, Universit\'e Paris-Saclay, CEA Saclay, 91191 Gif-sur-Yvette Cedex, France.}
\affil[2]{Lawrence Berkeley National Laboratory, Energy Storage Group, 1 Cyclotron Road, Berkeley, CA 94720, USA.}
\affil[3]{CEA/DRT, LITEN, DTNM, 38054 Grenoble, France.}
\affil[*]{corresponding author: jmzanotti@cea.fr}

\keywords{Ionic Liquids, conductivity, carbon nanotube membrane, Battery}
\twocolumn

\begin{document}

\flushbottom
\maketitle
\noindent\textbf{Ionic Liquids (ILs) are organic molten salts characterized by the total absence of solvent. They show remarkable properties: low vapor pressure, high ionic conductivity, high chemical, thermal and electrochemical stability. \cite{zhang2016nanoconfined} These electrolytes meet therefore key criteria for the development of safe energy storage systems. Due to a competition between electrostatic and van der Walls interactions, ILs show an uncommon property for neat bulk liquids: they self-organize in transient nanometric domains. \cite{hayes_structure_2015} In ILs-based electrochemical devices, this fluctuating nano-segregation acts as energy barriers to the long range diffusional processes and hence to the ionic conductivity.  \cite{berrod2017sci-rep} Here, we show how the ionic conductivity of ILs can be increased by more than one order of magnitude by exploiting one dimensional (1D) confinement effects in macroscopically oriented carbon nanotube (CNT) membranes. We identify 1D CNT membranes as promising separators for high instant power batteries.}

\begin{figure*}[h!]
\centering
\includegraphics[width=.85\textwidth]{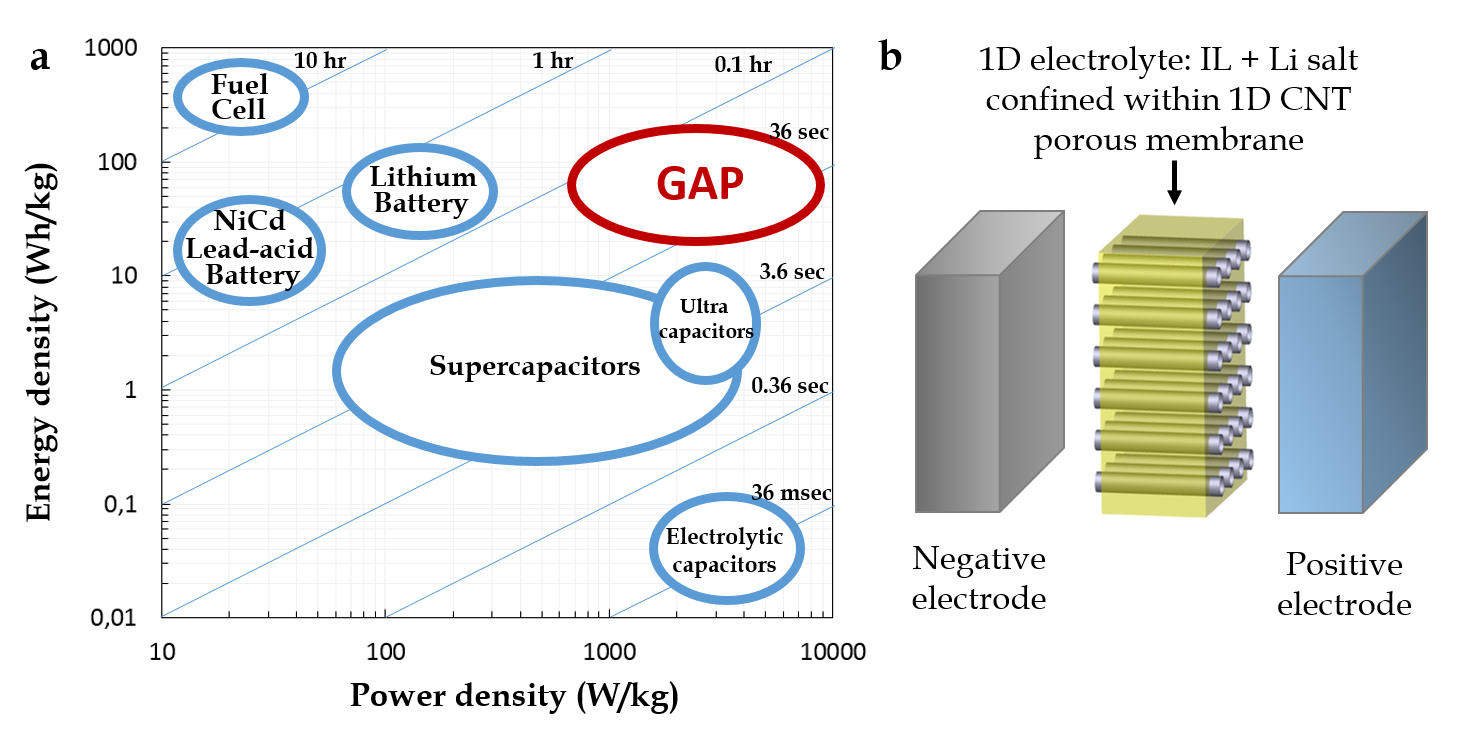}
\caption{\textbf{Lithium batteries: the quest for high power, and the 1D CNT battery separator.} \textbf{a}, Ragone plot highlighting the performance of different electro-chemical systems. The diagonal solid lines indicate typical charge/discharge duration of the different devices. In batteries, this time is partly driven by the electrolyte transport properties within the separator. Enhancing the transport properties of the electrolyte is a direct way to increase the power density of batteries. \textbf{b}, In bulk, the spontaneous nanometric self-organization of ILs induces large density fluctuations that represent a detrimental condition for ionic conductivity. Nanometric confinement of an IL within a volume smaller than the characteristic size of the spontaneous fluctuations is an efficient way to frustrate this phenomenon. To improve the overall conductivity this confinement should take place within pores showing no tortuosity. The alliance of \textit{i)} the frustration of the the confined IL nanostructuration, and \textit{ii)} no tortuosity, is a way to enhance the instant power of batteries and fill-in the gap shown in a. From patent Berrod \textit{et al.} \cite{berrod_nanocomposite_2016}. }
\label{Fig_1}
\end{figure*}

Water confined inside a 1.4 nm diameter CNT is a perfect example of a possible breathtaking effect induced by extreme nanometric confinement: to accommodate such a tight space, the molecules adopt a core-shell organization. \cite{kolesnikov_anomalously_2004} This very specific molecular structuration, controlled by the CNT internal curvature, \cite{falk_molecular_2010} induces consequences for the transport properties at a scale orders of magnitude larger: at the macroscopic scale, when forced inside a CNT membrane, water is found to flow-up 2-3 orders of magnitude faster than predicted by the continuum hydrodynamics picture. \cite{holt_fast_2006} 
Secchi \textit{et al.} also report on a significantly enhanced permeability of water inside a single CNT, while no modification is observed in a Boron Nitride nanotube (BNNT) of similar diameter.\cite{secchi2016massiveflow, secchi2016singleCNT}  These both types of nanotubes are crystallographically similar, but electronically different. CNT, (being semi-metallic)  induce an hydrodynamic slippage of water at the carbon surface of hundreds of nm, while no such effect is observed in BNNT (being insulating), highlighting the role of the electronic structure of the confining material. This gigantic mass flow property, referred as superlubricity, is now proposed for industrial outcomes in the field of water treatment.\cite{lee_are_2015}  
This exciting framework, leads us to propose to extend the results obtained on 1D confined water to the case of electrolytes to boost their transport properties.\cite{berrod_enhanced_2016,berrod_nanocomposite_2016} Indeed, electrochemical devices such as batteries designed for electric vehicles must provide both high energy and power densities. In comparison to other electrochemical systems, lithium batteries deliver very high energy densities (150 Wh/kg). \cite{tarascon_issues_2001} Supercapacitors provide relatively high power densities (10 kW/kg) but at the expense of the total energy they can store (5 Wh/kg). Up to date, no system fulfills both the power and energy requirements. Accordingly, a gap appears in the Ragone Power/Energy plot (Fig. \ref{Fig_1}a). Taking advantage of 1D gigantic transport properties, and thus ionic conductivity, would be a direct way to increase the power density of electrochemical devices.

For the last decades, tremendous efforts have been made to improve the performances of batteries. Better electrodes and electrolytes / separators have been developed but the power/energy gain is usually incremental.\cite{2015rev-li-batt}
Here, we show that 1D CNT based confinement matrix can increase the ILs ionic conductivity by more than one order of magnitude. Due to their noteworthy properties mentioned above, ILs are promising electrolytes for batteries. \cite{larcher_towards_2015} The combination of a large diversity of anions and cations lead to a flourishing variety of ILs,\cite{hayes_structure_2015} but they all share a limiting common ground: a quite high cost and the presence of a semi-local segregation in bulk (few nm \cite{triolo_nanoscale_2007} up to few tens of nm \cite{kirchner_ion_2015}). Here, using 1D nanometric confinement (Fig. \ref{Fig_1}b) we target to: \textit{i)} prevent the ILs spontaneous nanostructuration, \textit{ii)} avoid any tortuous pathway of lithium ions traveling between the electrodes, \textit{iii)} prevent friction at the CNT / ILs interface (smooth surface of the CNT) \cite{falk_molecular_2010,berrod_enhanced_2016} and \textit{iv)} get close to the superlubricity state, as observed in the case of water confined in CNT. \cite{kolesnikov_anomalously_2004}
%
%
We indeed show in this paper that a dramatical enhancement of the ionic conductivity is reached under confinement.

\begin{figure*}[h!]
\includegraphics[width=1\textwidth]{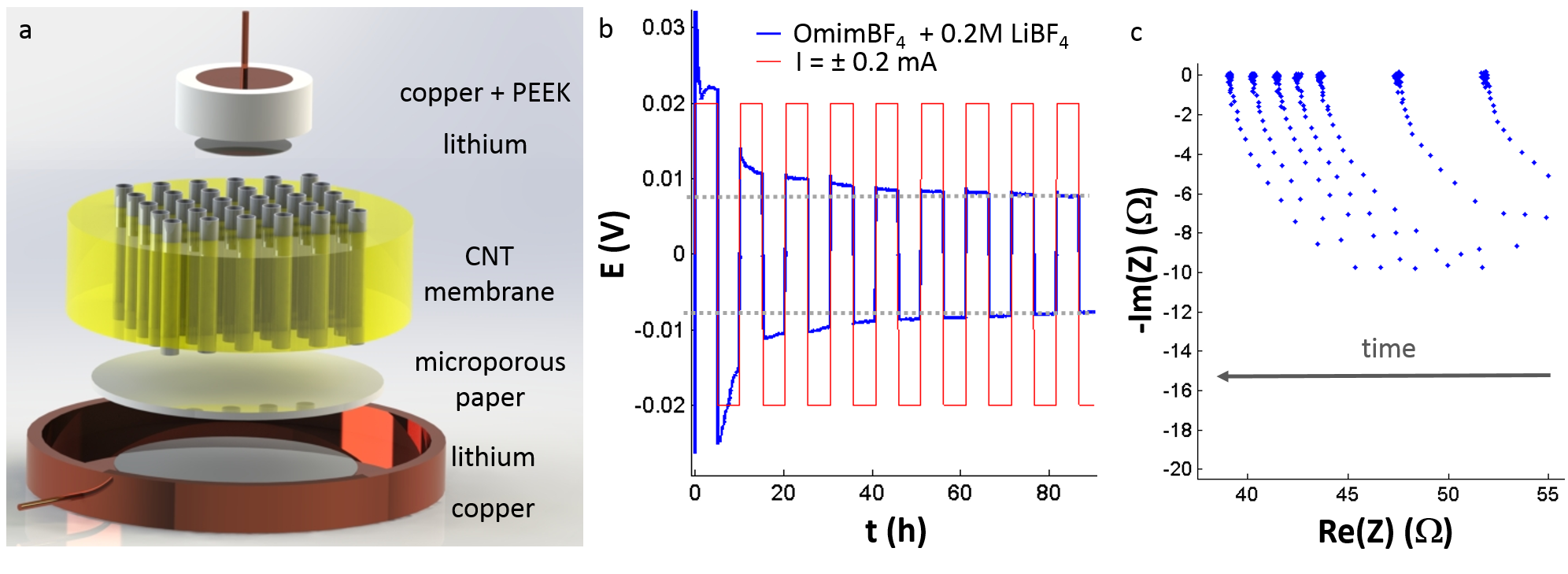}
\caption{\textbf{a.} Schematic view of the symmetric cell designed for the GC and EIS measurements (performed in a glove box under inert atmosphere). The CNT membrane (CNT: dark gray, polymer: yellow, typically 100 $\mu$m thick) is pressed in between two lithium foils (gray). To avoid any short circuit, a 155 $\mu$m thick microporous paper (light gray) is intercalated. The upper electrode is insulated with a polymer (PEEK, white). The electrolyte is added in excess in the copper reservoir. The CNT pores are spontaneously filled by capillarity and the electrolyte does not evaporate during the measurement. The whole cell is caught in a vice to apply a constant pressure on the samples. \textbf{b.} GC spectra obtained for OmimBF$_4$ + 0.2M LiBF$_4$ with a current of 0.2 mA. After 2 days, the total potential, $U_\text{tot}$, reaches a plateau around 7.7 mV, corresponding to a total impedance, $R_\text{tot}$, of 38.5 $\Omega$. Gray dotted lines are guides for the eyes to show the plateau value. \textbf{c.} Selected EIS spectra measured (from 1 to 10$^6$ Hz) between each cycles for the same sample. After few cycles, $R_\text{tot}$ stabilizes around 39 $\Omega$. Both EIS and GC measurements are in good agreement.}
\label{Fig_2}
\end{figure*}
We focus on two canonical imidazolium based ILs: 1-butyl-3-methylimidazolium bis trifluoromethanesulfonyl imide (BmimTFSI) and 1-methyl-3-octylimidazolium tetrafluoroborate (OmimBF$_4$). The electrolytes are prepared by adding lithium salts (LiTFSI and LiBF$_4$) at various concentration (0.2M up to 1M) to the neat ILs. The ionic conductivity of the ILs is determined by electrochemical impedance spectroscopy (EIS) and galvanostatic cycle (GC) measurements, in bulk and under CNT confinement. All the measurements have been performed in a glove box at 298 K using a specifically designed cell shown in Fig. \ref{Fig_2}a.\cite{landesfeind2016tortuosity}
The CNT membrane has been prepared following a process we already published.\cite{berrod_enhanced_2016} It is then filled with the electrolyte and pressed in between two lithium foils. To avoid any short circuit, a 155 $\mu$m thick microporous paper is used as an electrical insulator. We note that the CNT electronic conductivity is unacceptable for a battery separator. However, several strategies such as using semi-conductors CNT, or grafting polymer onto the CNT can be employed to overcome this aspect. Here, we focus our study on the effect of extreme confinement on the ILs ionic conductivity, that could lead to a high power battery separator filling the gap shown in the Ragone plot (Fig.\ref{Fig_1}a).

%

GC are performed using a current of 0.2 mA (1 mA / cm$^2$ with a 0.2 cm$^2$ electrode) and a 1 to 10 hours charge / discharge cylces. Between each cycle, an EIS measurement is performed. Fig. \ref{Fig_2}b shows a GC obtained for OmimBF$_4$ + 0.2M LiBF$_4$. After 2 days, we observe that the potential stabilizes at 7.7 mV. This asymptotic value is used to determine the total impedance, $R_\text{tot}$, of the system (copper cell, insulator and electrolyte): $R_\text{tot} = U_\text{tot} / I$.
To determine the contribution of the copper cell and of the microporous paper, $R_\text{cell}$, the measurement is then repeated with no CNT membrane (Fig. \ref{Fig_S5}). The impedance arising from the IL confined in the CNT, $R_\text{@CNT}$, is obtained by subtracting $R_\text{cell}$ to the total impedance, $R_\text{tot}$: $R_\text{@CNT} = R_\text{tot} - R_\text{cell}$.
Given the value of the porosity and the thickness of the CNT membrane, one can calculate the ionic conductivity of the confined IL:  $	\sigma_\text{@CNT} = t / \left( R_\text{@CNT} \times S \right)$, where  $t$ and $S$ stand for the thickness of the CNT membrane (typically 100 $\mu$m), and the surface of the CNT in contact with the upper electrode, respectively (cf \textit{Methods} for details). 
Fig. \ref{Fig_2}c shows the EIS spectra recorded between each cycles. The high frequency value of the complex impedance corresponds to the total impedance of the system. The impedance measured is in good agreement with the one determined by GC: we find $R_\text{tot} =$ 38.5 $ \pm 1 ~\Omega$ and 39 $\pm 1 ~\Omega$ using GC and EIS, respectively (Fig. \ref{Fig_2}).
The GC and EIS spectra obtained for all the electrolytes studied here are shown in Fig.\ref{Fig_SI_bk}-\ref{Fig_S4}. The conductivity values of bulk and confined ILs are given in Table \ref{table_1}, together with the ionic conductivity gain, $\sigma_\text{gain}$, defined as: $\sigma_\text{gain} = \sigma_\text{@CNT} / \sigma_\text{bulk}$. We need to stress that $R_\text{cell}$ is usually quite small compared to $R_\text{@CNT}$ ($R_\text{cell}$ $<$ 10\% of the total impedance), which drastically reduces the experimental error on the conductivity gain determination (Fig. \ref{Fig_S5}). In all cases, we observe that CNT confinement induces a significant increase (up to a factor 56) of the ionic conductivity (Table \ref{table_1}).

\begin{table}[h!]
\centering
\begin{tabular}{>{\columncolor{maroon!10}}lc|c|c}
\arrayrulecolor{maroon!30} 
\rowcolor{maroon!30}  & \bf{$\sigma_\textbf{bulk}$ }  & \bf{$\sigma_\textbf{@CNT}$}   & \bf{$\sigma_\textbf{gain}$ } \\
\hline
\arrayrulecolor{white} 
\rowcolor{maroon!30} \bf{Electrolyte} & \bf{ (S/m)}  & \bf{(S/m)}   &  \\
\hline
 BmimTFSI + & \multirow{2}{*}{0.34$\pm$0.03} & \multirow{2}{*}{2.6 $\pm$ 0.5} & \multirow{2}{*}{7.6 $\pm$ 2.0}\\
 0.2M LiTFSI & & & \\
\hline
BmimTFSI + & \multirow{2}{*}{0.14$\pm$0.01} & \multirow{2}{*}{1.7 $\pm$ 0.3} & \multirow{2}{*}{12 $\pm$ 3}\\
1M LiTFSI & & & \\
\hline
OmimBF$_4 $ + & \multirow{2}{*}{0.061$\pm$0.006} & \multirow{2}{*}{3.4 $\pm$ 0.6} & \multirow{2}{*}{56 $\pm$ 17}\\
0.2M LiBF$_4$ & & & \\
\hline
OmimBF$_4 $ + & \multirow{2}{*}{0.032$\pm$0.003} & \multirow{2}{*}{0.56$\pm$0.11} & \multirow{2}{*}{17 $\pm$ 5}\\
1M LiBF$_4$ & & & \\
\hline
OmimBF$_4 $ + & \multirow{2}{*}{0.057$\pm$0.006} & \multirow{2}{*}{1.6 $\pm$ 0.3} & \multirow{2}{*}{28 $\pm$ 8}\\
0.85M LiTFSI & & & \\
					
\end{tabular}
\caption{Electrolytes conductivity in bulk and under confinement within the CNT. The conductivity gain ($\sigma_\text{gain} = \sigma_\text{@CNT} / \sigma_\text{Bulk}$) is given.}
\label{table_1}
\end{table}
In bulk,  a characteristic feature of ILs is the spontaneous nano-segregation process observed experimentally by diffraction with the appearance of the so-called pre-peak in the [0.2-0.4] \AA$^{-1}$ region. The phenomenon is mainly controlled by the alkyl side-chain length, $n_C$, carried by the imidazolium ring.\cite{martinelli_insights_2013}  The anion also plays a role in the segregation, but only at a second order. The longer the alkyl chain, the stronger the nanostructuration. It has recently been evidenced that, important detrimental consequences on the ionic conductivity arise from the ILs nanostructuration and the related viscosity fluctuations. \cite{berrod2017sci-rep, ferdeghini2017nanostructuration}  It induces transient energy barriers disrupting the Fickian process theoretically driving the charges transport from one electrode to the other.

In our view, next to this situation in bulk,  the conductivity gains we observe here (Table \ref{table_1}) can be understood in light of recent progresses on the physics of ILs under nanometric confinement. Specific ILs structural rearrangements induced by confinement in CNT have been reported.\cite{chen_morphology_2009, pensado2014CNT-IL}
It has also been observed that ionic conductivity of BmimTFSI through 1D single nanopore leads to a one order of magnitude gain in conductivity.\cite{tasserit_pink_2010} Detailed analysis of these voltage-clamp results show that, upon confinement, the transport of charge carriers in confined BmimTFSI ($n_C$ = 4) is strongly facilitated, while the regular bulk behavior is observed in the case of the confined EmimSCN ($n_C$ = 2). Furthermore, we have shown that the self-diffusion coefficient of neat OmimBF$_4$ ($n_C$ = 8) can be increased by a factor 3 upon confinement in the same 1D CNT membranes (4 nm diameter) while no gain was found for BmimTFSI ($n_C$ = 4).\cite{berrod_enhanced_2016}
Finally, recent MD results indicate that imposing confinement of ILs in CNT with diameter under 2 nm exalts the transport properties up to several orders of magnitude, highlighting the crucial role of the pore diameter.\cite{ghoufi_ultrafast_2016,chaban2014MD-CNT} 
Here, depending on the IL cation (Omim or Bmim), the IL anion (TFSI or BF$_4$), and on the lithium salt nature (LiTFSI or LiBF$_4$) and concentration (0.2, 0.85 and 1 M), at the present stage of our study, it is difficult to delineate a simple trend on the evolution of $\sigma_\text{gain}$. A firm conclusion can nevertheless be reached: compared to their bulk analogues, for all of the five systems investigated here, very significant conductivity gains (from 7.6 to 56) are observed when they are in situation of 1D CNT confinement.
Our interpretation, is that this conductivity gain is a direct visible consequence of a strong structural change induced by the confinement. In particular, the ionic conductivity gain is higher for OmimBF$_4$ ($n_C$ = 8 showing a strong pre-peak) than BmimTFSI ($n_C$ = 4 having a weaker nanostructuration), which is in good agreement with the literature results described above. %
Either,  the spontaneous formation of the transient aggregates is strongly frustrated when the fluid is forced in channels with a diameter smaller, or close to the bulk nano-segregation characteristic size, either the confinement modifies the molecules configurations for example by inducing specific structures favoring the lithium ions pathway (preferential orientation in the direction of the pore axis such as concentric cylinders).\cite{pensado2014CNT-IL} In such a situation, one can envision that at the molecular level the dynamics of at least a fraction of the IL and/or counter ions is frustrated. This could lead to a decoupling of the lithium ion dynamics from the surrounding media and result in a strong increase of the sole lithium ion transport number.  

To conclude on a more technological perspective, we note that fully optimizing complex devices as batteries systems comes to finely investigate the whole chain of structural/kinetic processes: electrodes morphological evolution upon lithiation, lithiation kinetics and specific chemical reactions at the electrode/electrolyte interface.\cite{liu_pomegranate-inspired_2014} Thanks to an intense worldwide scientific and technological activity, constant and regular progresses are obtained by improving the capabilities of each element of batteries: electrodes, high voltage electrolytes and electrode/electrolyte interface. But these incremental step by step improvements, of the order of 10\% each time, are far from being able to challenge the logarithmic scale gap shown in red on the Ragone plot (Fig. \ref{Fig_1}a). 
Here, we provide evidences that, with respect to the bulk situation, confinement of ILs within 1D macroscopically oriented nanometric channels, leads to a drastic enhancement of the ionic conductivity (up to a factor 56). 1D CNT membranes appear therefore to be a promising battery separator.
Next to the fundamental interest, 1D CNT forests are also, appealing complementary elements for batteries: electrodes showing large specific surface, potentially high loading (several hundreds $\mu$m thick) and promoting fast 1D electronic conduction to the current collector. \cite{simon_capacitive_2013} The expected significant gains in 1D electronic/ionic conductivity in both  the electrodes/separator respectively should benefit to the instant power (or charge time) of the assembly. Boosting this property which is a notorious severe limitation of the electrochemical devices developed so far, could turn-out to be a key asset to favor the development of low carbon transportation alternatives (electric/hybrid vehicles).

\begin{small}
\bibliography{Biblio_paper}   

\section*{Acknowledgments}
Q.B., P.J., A.F, J.D. and J.-M.Z. thank the \textit{Programme CEA transverse NTE} for funding. Q.B. was supported by an Outgoing CEA fellowship from the CEA-Enhanced Eurotalents program, co-funded by FP7 Marie-sk\l odowska-Curie COFUND program (Grant Agreement 60000382). We thank P. Lavie for the drawings, and B. Homatter for the experimental cells manufacturing.

\section*{Author contributions}
Q.B. prepared the samples and performed the EIS and GC measurements, P.J. , V.B. and Y.F. supervise the electrochemical measurements, J.-M.Z. initiated and supervised the project, A.F. and J.D. developed and synthesized the CNT carpets. All the authors contributed to the preparation of the manuscript.


\section*{Additional information}
Correspondence and requests for materials should be addressed to J.-M.Z.

\section*{Competing financial interests}
The authors declare no competing financial interests.

\newpage

\section*{Methods}
\label{Methods}
All the measurements have been performed in a glove box at 298 K. 

\paragraph{Electrolytes preparation.}
Before being introduced inside the glove box, lithium salts and ILs have been dried for 1 night under vacuum at 120$\degree$C. LiBF$_4$ (EM Science) and LiTFSI (99.95 \% purity, sigma aldrich) salts have been added to neat ILs (OmimBF$_4$ $>$ 99 \% purity, sigma aldrich ; and BmimTFSI $>$ 99 \% purity, solvionic) in glove box. Electrolytes are then stirred for few days at ambient temperature before use. The lithium salt concentration, C$_\text{Li}$, is determined as followed:
\begin{equation*}
C_\text{Li} =  \frac{m_\text{Li}}{M_\text{Li}} . \frac{1}{\frac{m_\text{Li}}{\rho_\text{Li}} + V_\text{IL}}
\end{equation*}
where $m_\text{Li}$, $M_\text{Li}$, $\rho_\text{Li}$, and $V_\text{IL}$ stand for the mass, the molar mass and the density of lithium salt; and the IL volume, respectively.

\paragraph{CNT membrane.}
\label{methods_CNT}
The CNT membranes has been prepared as explained in ref. \cite{berrod_enhanced_2016} The CNT density, $n_\text{CNT}$, is $3 \times 10^{11} \text{cnt/cm}^2$, the upper electrode is 0.5 cm in diameter, and the CNT internal diameter is 4 nm. The surface of the CNT in contact with the upper electrode, S, is therefore determined as followed:\\
\begin{equation*}
S = S_\text{elec} \times S_\text{CNT} \times n_\text{CNT}
\label{SI_eq1}
\end{equation*}
where $S_\text{elec}$ and $S_\text{CNT}$ stand for the surface of the upper electrode and one CNT, respectively. Then,
\begin{equation*}
S =  \pi .(0.5/2)^2 \times \pi \left(2.5\times10^{-9}\right)^2 \times 3.10^{11} = 0.7402 \times 10^{-6} ~\text{m}^2
\end{equation*}
As used here, the CNT membrane is not competitive for a battery since it is quite thick (100 $\mu$m) and having a porosity of only 6$\%$. The porosity and thickness have been chosen for experimental convenience. However, it has been shown that CNT forest can be mechanically densified \cite{lee_carbon_2015,jeon2015dens}, and the CNT growth is easily tunable and the CNT can be few $\mu$m to few hundreds $\mu$m high.

\paragraph{Electrochemical measurements}
\label{methods_EIS_GC}%
EIS and GC measurements have been performed on a Bio-Logic VMP3 instrument, with, a frequency range of [1 - 10$^6$] Hz and a current of 0.2 mA (corresponding to 1 mA / cm$^2$), respectively.

\end{small}


\cleardoublepage
\newpage
~
\newpage

\onecolumn

\noindent\textbf{\large Supplementary Information}
\vspace{0.5 cm}

\begin{center}
\textbf{\huge Ionic liquids confined in 1D CNT membranes: gigantic ionic conductivity}

\vspace{0.5 cm}
\textbf{Q. Berrod, P. Judeinstein, Y. Fu, V.S. Battaglia, A.  Fournier, J. Dijon, J.-M. Zanotti}
\end{center}

\vspace{0.5 cm}

\renewcommand{\theequation}{S\arabic{equation}}
\renewcommand{\thefigure}{S\arabic{figure}}

\setcounter{section}{0} 
\setcounter{equation}{0}
\setcounter{figure}{0}
\setcounter{table}{0}
\setcounter{page}{1}

\section{EIS and GC measurements}
\label{SI_EIS}

\subsection{Bulk ILs}
\begin{figure*}[h!]
\centering
\includegraphics[width=.8\textwidth]{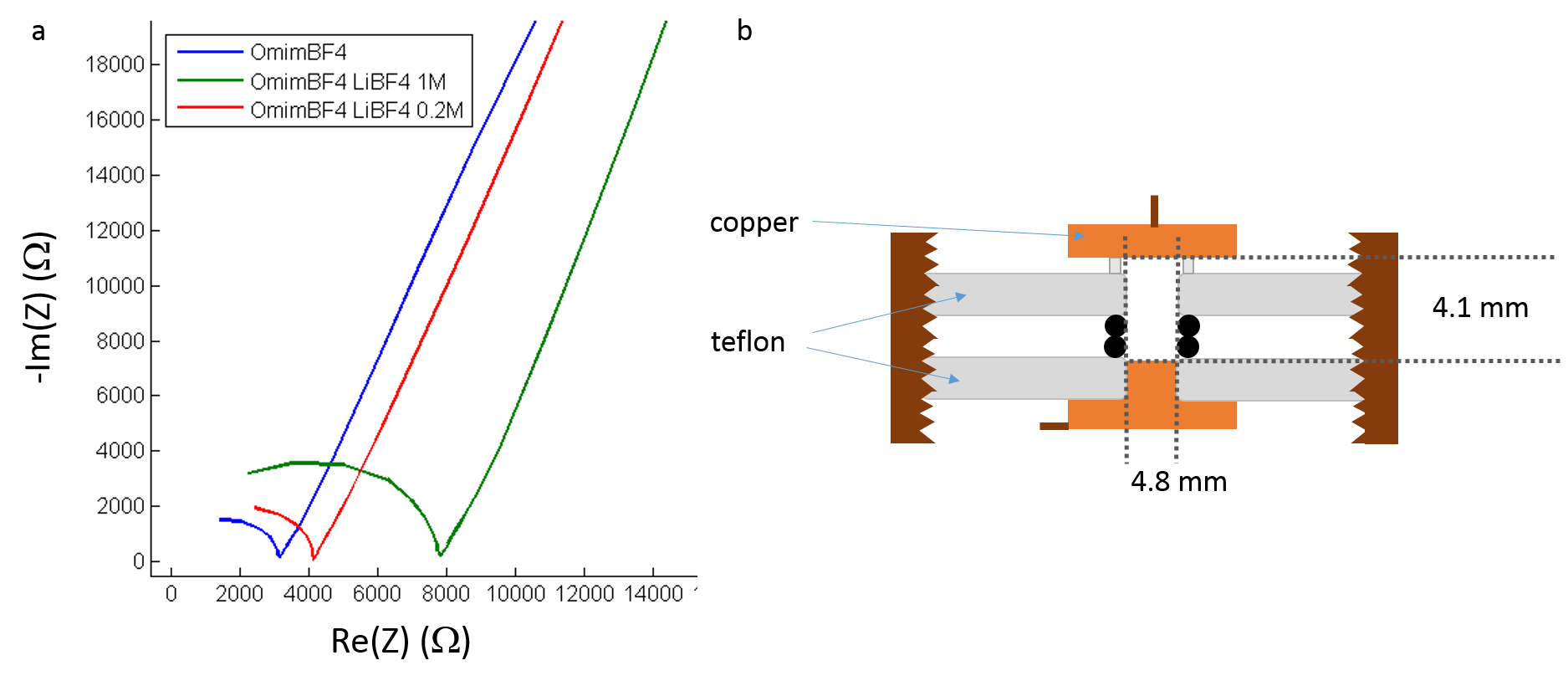}
\caption{\textbf{a.} EIS spectra measured for OmimBF$_4$, OmimBF$_4$ + 0.2M LiBF$_4$, and OmimBF$_4$ + 1M LiBF$_4$. \textbf{b.} Cross section of the cylindrical cell designed for the EIS measurements of the bulk ILs. }
\label{Fig_SI_bk}
\end{figure*}

EIS measurements have been performed in the following experimental cell (Fig.\ref{Fig_SI_bk}b). A typical EIS spectra is shown in Fig.\ref{Fig_SI_bk}a.
 
\subsection{OmimBF$_4$ + 1M LiBF$_4$}

\begin{figure*}[h!]
\centering
\includegraphics[width=.9\textwidth]{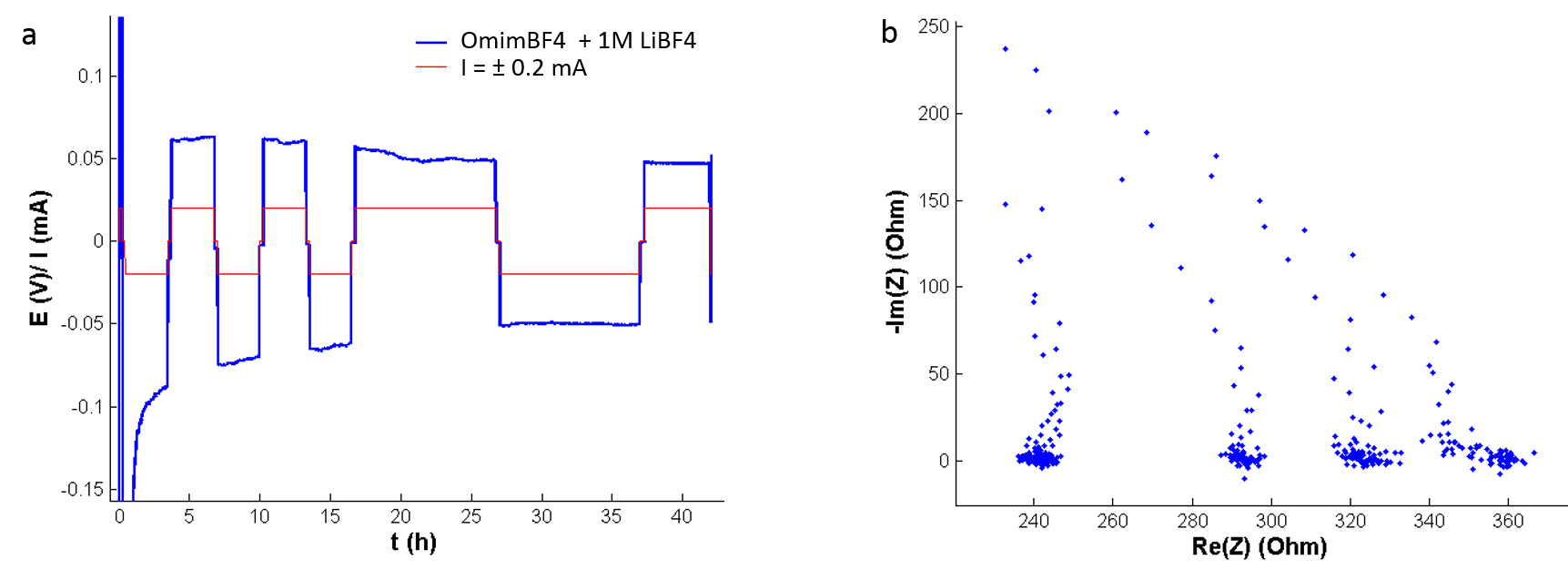}
\caption{\textbf{a.} GC spectra obtained for OmimBF$_4$ + 1M LiBF$_4$ with a current of 0.2 mA. The total potential, $U_\text{tot}$, reaches a plateau around 47 mV, corresponding to a total impedance, $R_\text{tot}$, of 235 $\Omega$. \textbf{b.} EIS spectra measured between each cycles for the same sample. After few cycles, $R_\text{tot}$ stabilize around 240 $\Omega$. Both EIS and GC measurements are in good agreement.}
\label{Fig_S2}
\end{figure*}

\newpage

\subsection{BmimTFSI + LiTFSI (0.2M \& 1M)}

\begin{figure*}[h!]
\centering
\includegraphics[width=.95\textwidth]{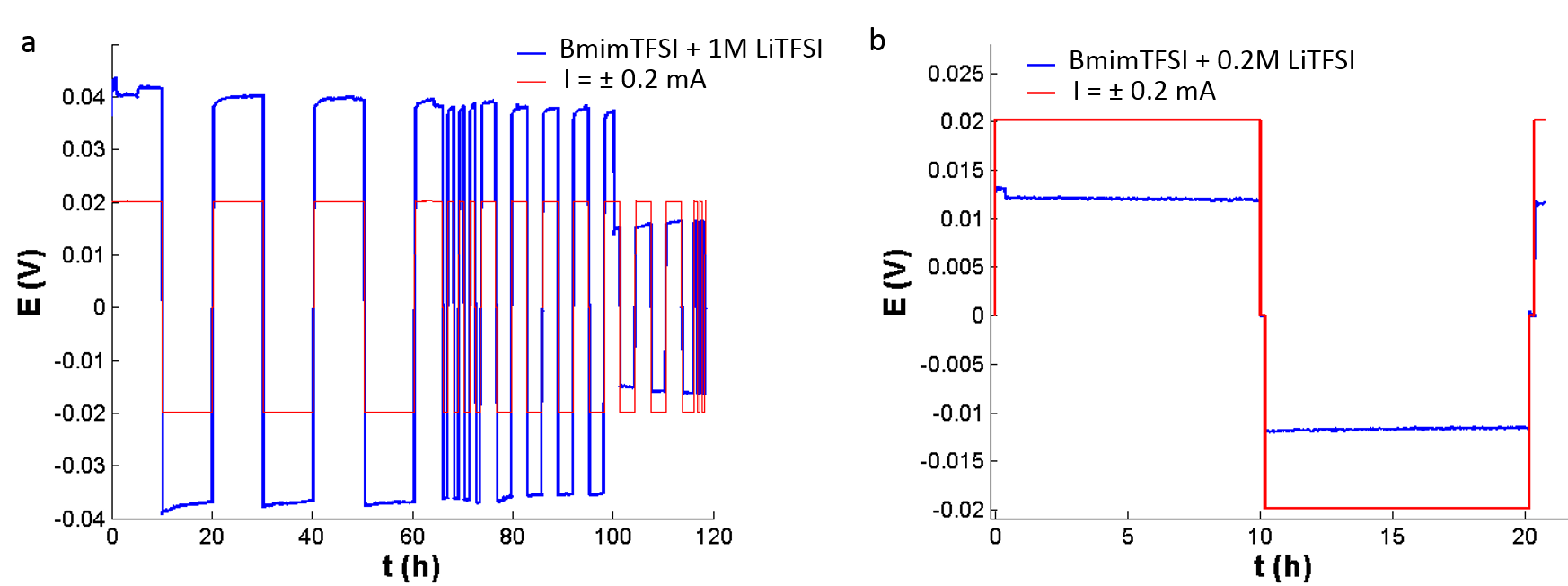}
\caption{\textbf{a.} GC spectra obtained for BmimTFSI + 1M LiTFSI with a current of 0.2 mA. The total potential, $U_\text{tot}$, reaches a plateau around 16 mV, corresponding to a total impedance, $R_\text{tot}$, of 80 $\Omega$. \textbf{b.} GC spectra obtained for BmimTFSI + 1M LiTFSI with a current of 0.2 mA. The total potential, $U_\text{tot}$, reaches a plateau around 12 mV, corresponding to a total impedance, $R_\text{tot}$, of 60 $\Omega$. }
\label{Fig_S3}
\end{figure*}

\subsection{OmimBF$_4$ + 0.85M LiTFSI}

\begin{figure*}[h!]
\includegraphics[width=.5\textwidth]{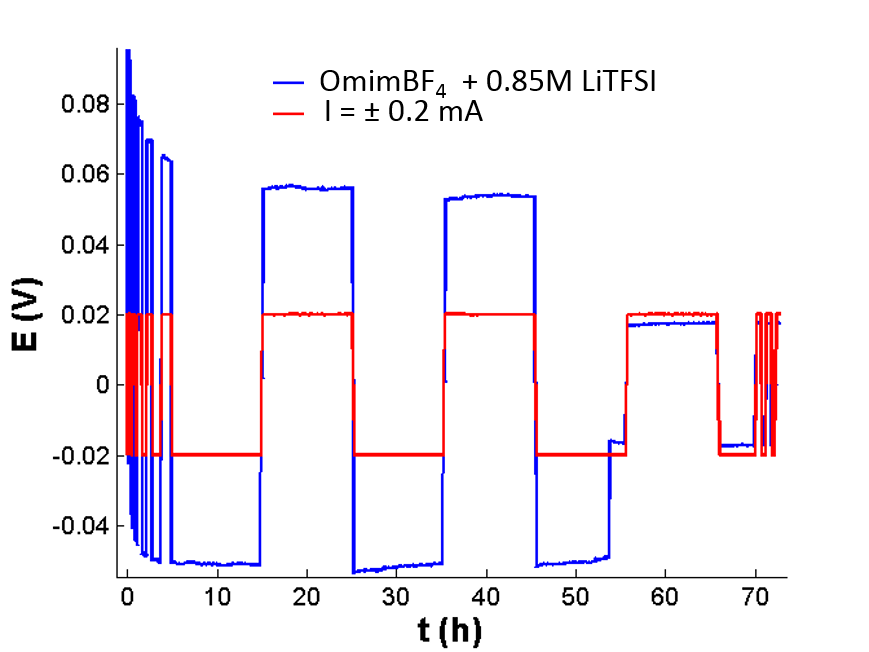}
\caption{GC spectra obtained for OmimBF$_4$ + 0.85M LiTFSI with a current of 0.2 mA. The total potential, $U_\text{tot}$, reaches a plateau around 17 mV, corresponding to a total impedance, $R_\text{tot}$, of 85 $\Omega$. }
\label{Fig_S4}
\end{figure*}

\newpage
\subsection{Contribution of the experimental cell}

\begin{figure*}[h!]
\includegraphics[width=.95\textwidth]{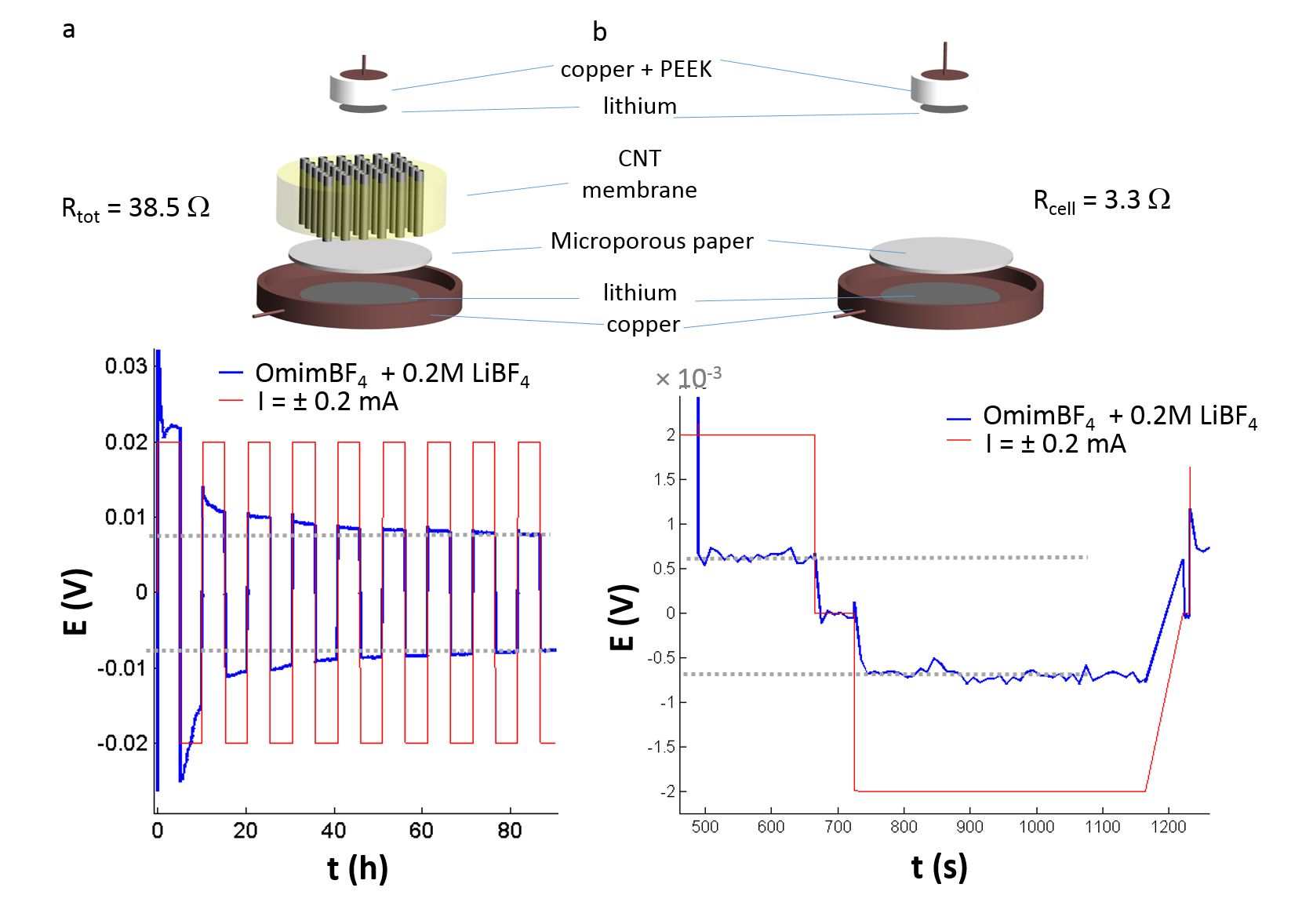}
\caption{\textbf{a.}. GC spectra obtained for OmimBF$_4$ + 0.2M LiBF$_4$ with a current of 0.2 mA. The total potential, $U_\text{tot}$, reaches a plateau around 7.7 mV, corresponding to a total impedance, $R_\text{tot}$, of 38.5 $\Omega$. \textbf{b.}. GC spectra obtained for OmimBF$_4$ + 0.2M LiBF$_4$ with a current of 0.2 mA, and without the CNT membrane. The total potential, $U_\text{cell}$, reaches a plateau around 0.65 mV, corresponding to a total impedance, $R_\text{cell}$, of 3.3 $\Omega$. Gray dotted lines are guides for the eyes to show the plateau value.}
\label{Fig_S5}
\end{figure*}

Typical GC spectra with and without the CNT membrane are shown in Fig.\ref{Fig_S5} a and b, respectively. The total impedance, $R_{\text{tot}}$ is much higher with the CNT membrane than $R_\text{cell}$, the impedance arising from the cell, both electrodes, and the microporous filter ; due the low CNT membrane porosity. Note that as $R_\text{cell}$ is roughly 10\% of the total impedance. The error that can be done on the determination of the impedance arising from the CNT membrane, $R_{\text{@CNT}}$, is therefore limited ($R_\text{@CNT} = R_\text{tot} - R_\text{cell}$).

\end{document}